\documentclass{jpsj3}
\usepackage{txfonts}
\usepackage{color}
\title{Estimate of the Critical Exponent of the Anderson Transition in the Three and Four Dimensional Unitary Universality Classes}

\author{Keith Slevin$^1$\thanks{slevin@phys.sci.osaka-u.ac.jp} and Tomi Ohtsuki$^2$}
\inst{
$^1$Department of Physics, Graduate School of Science, Osaka University, 1-1 Machikaneyama, Toyonaka, Osaka 560-0043, Japan\\
$^2$Division of Physics, Sophia University, Chiyoda-ku, Tokyo 102-8554, Japan} 

\abst{
Disordered non-interacting systems are classified into ten symmetry classes, with the unitary class
being the most fundamental.
The three and four dimensional unitary universality classes are attracting renewed interest because
of their relation to three dimensional Weyl semi-metals and four dimensional topological insulators.
Determining the critical exponent of the correlation/localistion length for the Anderson transition in these classes is important both theoretically and experimentally.
Using the transfer matrix technique, we report numerical estimations of the critical exponent in a U(1) model in three and four dimensions.}

\begin{document}
\maketitle

\section{Introduction}
More than half a century after its discovery\cite{Anderson58}, the Anderson transition continues to attract attention.
The recent experimental realisation\cite{chabe08} in a cold atom system of the quantum kicked rotor (a system which has been mapped to the Anderson localisation problem\cite{Grempel84,haakebook}) has opened a new avenue for research of Anderson localisation and the Anderson transition.
The measurement\cite{lopez12} of the critical exponent of the correlation/localisation length for the Anderson transition in this experiment is in very good agreement with high accuracy numerical analysis for the three dimensional (3D) orthogonal universality class\cite{slevin99,Slevin14}.

Disordered non-interacting systems are classified into ten symmetry classes\cite{Zirnbauer96}.
The same symmetry classification\cite{Schnyder08,Kitaev09} is also applicable to topological insulators\cite{Hasan10,Qi11}.
Amongst these ten symmetry classes, the unitary class (class A) is the most general, where no symmetries such as time reversal, spin-rotation, chiral or particle-hole symmetries are present.
Systems in magnetic fields and systems with magnetic impurities, for example, belong to the unitary class.
The critical behavior of the Anderson transition in 3D systems subject to uniform magnetic fields was recently studied with high precision\cite{Ujfalusi15}.
Studies of disordered 3D systems subject to random magnetic fields\cite{Kawarabayashi98}
and with classical magnetic impurities\cite{Jung16} have also been reported.

In the quantum kicked rotor system, the effective dimensionality is determined by the
number of incommensurate frequencies used to modulate the amplitude of the periodic kick\cite{Casati89}.
The 3D orthogonal universality class has already been realised experimentally in this system\cite{chabe08}.
The addition of modulation by an extra incommensurate frequency would realise the 4D orthogonal universality class.
In addition, a version of the quantum kicked rotor in which the quantum Hall effect might be
realised has also been proposed\cite{Dahlhaus11}.
Thus, it is plausible that the 4D unitary universality class might be realised in a suitable
quantum kicked rotor system.
This universality class is also of interest from the view point of three dimensional Weyl semi-metals \cite{Zhang16}.
Just as two dimensional Dirac electrons appear on the surface of 3D topological
insulators, three dimensional Weyl electrons appear on the ``surface'' of 4D topological insulators
belonging to the unitary symmetry class.

A quantum Hall transition (QHT) has been predicted in 4D systems
\cite{Zhang01,Karabali02,Li13,Kraus13,Price15}.
On the one hand, in 2D, the unitary class exhibits no Anderson transition, while a QHT is observed.
On the other hand, the Anderson transition in the 2D symplectic class and the quantum spin Hall transition show the same critical exponent \cite{Obuse07}.
Whether the critical exponent of the
4D quantum Hall transition is the same as that of the Anderson transition in the 4D unitary class is an open problem.

In this paper, we study the 3D and 4D unitary universality classes and estimate the critical exponents with high precision
by studying numerically the U(1) model.  In the following section, we introduce the model,
and the method is explained in Section \ref{sec:tmm}.  The results of the finite size scaling analyses are shown in Section \ref{sec:fss},
followed by the discussion.

\section{Model}
\label{sec:model}
The U(1) model is a variant of Anderson's model of localisation\cite{Anderson58} in which time reversal symmetry is broken by
multiplying the unit hopping elements of that model by complex phases.
For the purpose of estimating critical exponents it is helpful that the length scale associated with
the breaking of time reversal symmetry is as short as possible.
This is achieved by using completely random phases\cite{Kawarabayashi98}.
The Hamiltonian of the U(1) model is
\begin{equation}\label{eq:hamiltonian}
  H = \sum_{i} E_i \left| i \right> \left< i \right|  - \sum_{\left<ij\right>} \exp\left(i\phi_{ij}\right) \left| i \right> \left< j \right| \;.
\end{equation}
Here, $\left| i \right>$ is an electron orbital centered on site $i$ of a $d$-dimensional simple cubic lattice
and the first sum is over all the sites of this lattice.
The lattice constant (taken as unity) sets the unit of length.
The second sum is over pairs of nearest neighbour sites on this lattice.

The magnitude of the hopping elements (taken as unity) between nearest neighbour sites on the lattice sets the unit of energy.
The orbital energies $E_i$ are independently and identically distributed according to the distribution
\begin{equation}
  P\left(E_i\right) = p\left(E_i\right) d E_i \;,
\end{equation}
where
\begin{equation}\label{eq:distribution}
p\left( {{E_i}} \right) = \left\{ {\begin{array}{*{20}{c}}
{{1 \mathord{\left/
 {\vphantom {1 W}} \right.
 \kern-\nulldelimiterspace} W}}&{\left| {{E_i}} \right| < {W \mathord{\left/
 {\vphantom {W 2}} \right.
 \kern-\nulldelimiterspace} 2}}\\
0&{{\rm{otherwise}}} \;.
\end{array}} \right.
\end{equation}
The scale of the fluctuations of the orbital energies is set by the parameter $W$.

As already mentioned the distribution of phases is taken as completely random, i.e. the $\phi_{ij}$ with
$i<j$ are independently and identically distributed uniformly on $[0,2\pi]$.
To ensure the Hamiltonian is Hermitian we set
\begin{equation}
  \phi_{ij} = -\phi_{ji} \;\;\; i>j \;.
\end{equation}

In the standard Wigner-Dyson classification\cite{Dyson62} the U(1) model has unitary symmetry, and in the more recent
classification\cite{Zirnbauer96} the model is in class A.
The ensemble of Hamiltonians described by Eq. (\ref{eq:hamiltonian})  is invariant under local U(1) gauge transformations.
For this reason we refer to the model as the U(1) model.

\section{Transfer matrix method}
\label{sec:tmm}
The brief description of the transfer matrix method that we give here follows closely
Ref. \citen{Slevin14}.
We refer the reader to that reference for further explanation and for details that are omitted.

In the transfer matrix method, the transmission of an electron with an arbitrary energy $E$
through a very long disordered wire is considered.
We denote the length of the wire by $L_x$ and consider cross sections $L\times L$ in the 3D
simulation and $L\times L \times L$ in the 4D simulation.
In practice, $L_x$ is many orders of magnitude larger than $L$.
We impose periodic boundary conditions on the wavefunction in the transverse directions.
Starting from the time independent Schr\"odinger equation with energy $E$ we derive a transfer
matrix product
\begin{equation}\label{eq:matrixproduct}
  M = \prod_{x=1}^{L_x} M_x \;.
\end{equation}
For sufficiently long disordered wires,
the transmission probability decreases exponentially with the length of the wire.\cite{Yuval74,Thouless77}
The associated exponential decay length is equal to the
reciprocal of the smallest positive Lyapunov exponent $\gamma$ of the matrix product Eq. (\ref{eq:matrixproduct}).
The Lyapunov exponents are the eigenvalues of the matrix
\begin{equation}
  \lim_{L_x \rightarrow \infty} \frac{\ln M^{\dagger} M}{2L_x} \;,
\end{equation}
and are estimated using the procedure described in detail in Ref. \citen{Slevin14}.

For the purpose of the transfer matrix calculations it is convenient to perform a gauge transformation of the
original Hamiltonian so as to eliminate all the phase factors appearing in hopping elements in the $x$-direction.
This transformation does not effect the values of the Lyapunov exponents.

\subsection{3D}
\begin{figure}
\includegraphics[width=15cm]{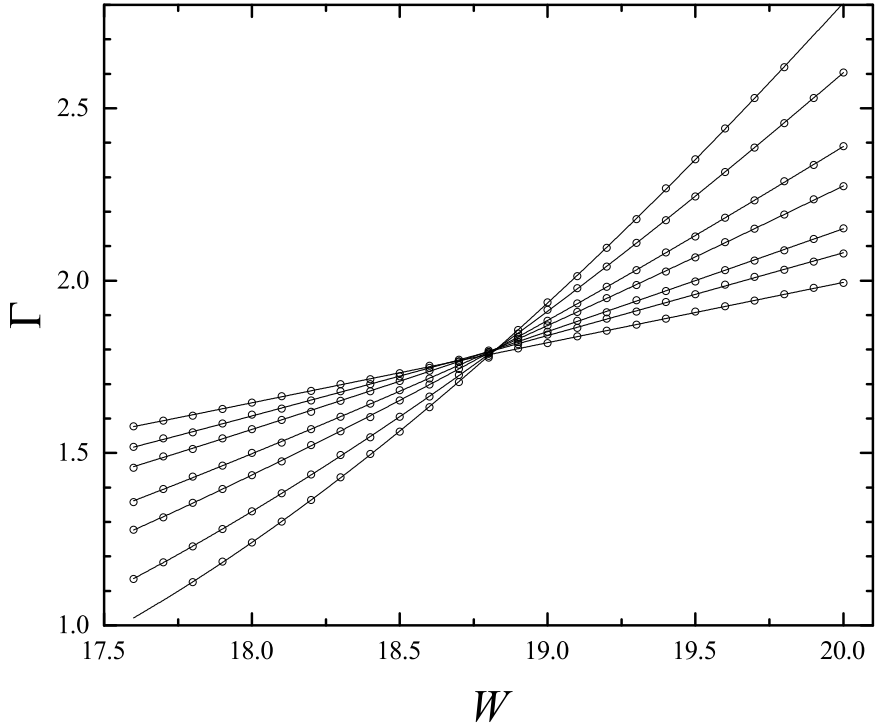}
\caption{Numerical data (circles) from which we estimated the critical parameters for the U(1) model in 3D.
The error in the data is less than the symbol size so we omit error bars.
We also show the finite size scaling fit (sold lines).}
\label{fig1}
\end{figure}

We set the energy $E$ at the band centre, i.e. $E=0$ and simulated the disorder range $17.6 \le W \le 20$ (except for the largest two system sizes where we used a slightly narrower range) for system sizes $L=4,6,8,12,16,24$ and $32$.
For each pair of $W$ and $L$, we estimated the smallest positive Lyapunov exponent $\gamma$ to a precision of $0.1\%$.
This required between approximately $L_x = 2\times 10^6$ and $3\times 10^7$ transfer matrix multiplications depending on the values of $W$ and $L$.
To avoid round off error we performed QR factorizations every $8$ transfer matrix multiplications and
to ensure the correct estimation of the precision we  aggregated the results of every 8 QR factorizations.
(In the notation of Secs. 2.3 and 2.4 of Ref. \citen{Slevin14}, we set $q=8$ and $r=8$.)
We show the data for this simulation in Fig. \ref{fig1}, where we plot the dimensionless quantity
\begin{equation}
  \Gamma = \gamma L \;,
\end{equation}
versus the disorder $W$.

\subsection{4D}
\begin{figure}
\includegraphics[width=15cm]{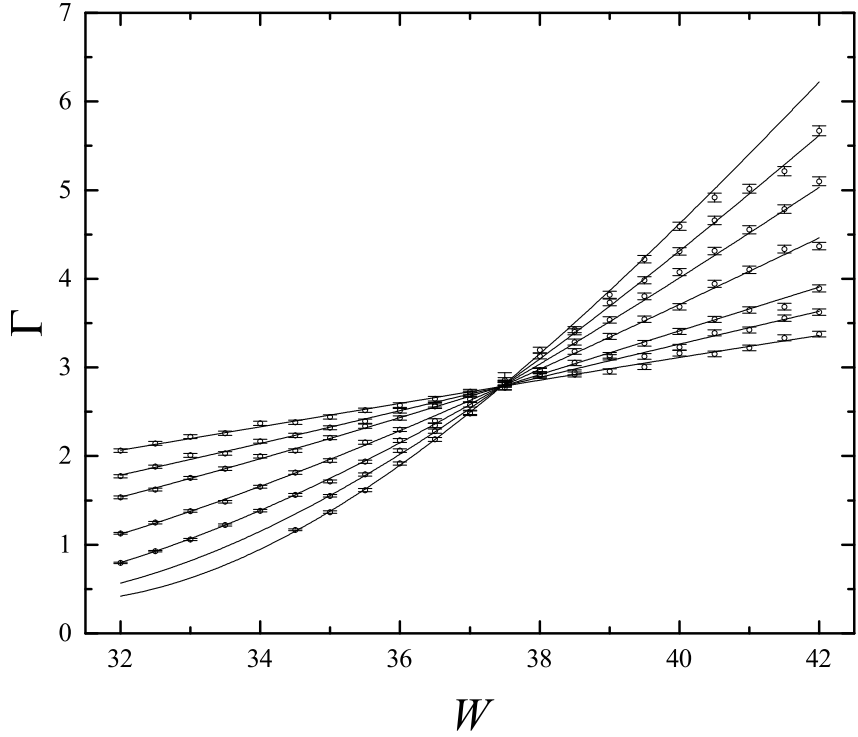}
\caption{Numerical data (circles), and their standard errors, from which we estimated the
critical parameters for the U(1) model in 4D.
We also show the finite size scaling fit (sold lines).}
\label{fig2}
\end{figure}

We set the energy $E$ at the band centre and simulated the disorder range $32 \le W \le 42$ (except for the largest two system sizes where we used a narrower range primarily because of the computational cost) for system sizes $L=4,6,8,12,16,20$ and $24$.
For each pair of $W$ and $L$, we estimated the smallest positive Lyapunov exponent $\gamma$ to a precision of $1.0\%$.
This required between approximately $L_x = 7\times 10^3$ and $2\times 10^5$ transfer matrix multiplications depending on the values of $W$ and $L$.
To avoid round off error we performed QR factorizations every $6$ transfer matrix multiplications and
to ensure the correct estimation of the precision we aggregated the results of every 12 QR factorizations.
We show the data for this simulation in Fig. \ref{fig2}.

\section{Finite size scaling analysis}
\label{sec:fss}
In our simulations we varied the disorder $W$ while keeping the energy $E$ fixed, so there is
a critical disorder $W_\mathrm{c}$, which is a function of energy, that separates the localised and extended phases.
At this critical disorder the localisation (correlation) length $\xi$ has a power law divergence
\begin{equation}
\label{eq:xiDivergence}
\xi \sim \left|W-W_\mathrm{c}\right|^{-\nu}\;,
\end{equation}
that is described by a critical exponent $\nu$.
We performed finite size scaling analyses to estimate the critical exponent $\nu$, the critical disorder $W_\mathrm{c}$
as well as other quantities (described below).
When corrections due to irrelevant variables could be neglected, either because the numerical
data had insufficiently high precision to resolve such corrections or because small system sizes had been excluded, we fitted the data to the equation
\begin{equation}
  \Gamma = F \left( \phi_1 \right) \;.
\end{equation}
When corrections due to irrelevant variables could not be neglected,
we fitted the data to the equation
\begin{equation}
  \Gamma = F \left( \phi_1, \phi_2 \right) \;.
\end{equation}
Here, $\phi_1$ and $\phi_2$ are scaling variables
\begin{equation}
  \phi_i = u_i L^{\alpha_i} \;, \;\; u_i = u_i\left( w \right) \;, \;\; i =1,2 \;,
\end{equation}
and
\begin{equation}
  w = W - W_\mathrm{c} \;.
\end{equation}
The first of these variables $\phi_1$ is a relevant scaling variable whose size dependence is related to the critical exponent $\nu$
\begin{equation}
 \nu = \frac{1}{\alpha_1} \;.
\end{equation}
The second of these variables is an irrelevant scaling variable that permits corrections to scaling due to irrelevant scaling variables to
be taken into account in an approximate manner.
The size dependence is described by an irrelevant exponent
\begin{equation}\label{eq:irrelevantexponent}
   y \equiv \alpha_2 < 0\;.
\end{equation}
The scaling variables are expanded in powers of $w$
\begin{equation}
\label{eq:nonlinearity}
  {u_i}\left( w \right) = \sum\limits_{j = 0}^{{m_i}} {{b_{i,j}}{w^j}} \;,
\end{equation}
which allows us to take account of possible nonlinearity of the scaling variables in the disorder $W$.
For the relevant scaling variable we must have
\begin{equation}\label{eq:u1}
  u_1\left( w = 0 \right) = 0 \;,
\end{equation}
so we fix $b_{1,0}=0$.
The scaling function is expanded in powers of the scaling variables
\begin{equation}\label{eq:scalingfunctionexpansion}
  F = \sum\limits_{{j_1} = 0}^{{n_1}} { \sum\limits_{{j_2} = 0}^{{n_2}} { {a_{{j_1}, {j_2} }}\phi _1^{{j_1}}  \phi _2^{{j_2}}  } }\;.
\end{equation}
To avoid ambiguity in the definition of the fit we fix
\begin{equation}\label{eq:firstordercoefficients}
  a_{1,0} = a_{0,1} = 1 \;.
\end{equation}
The constant term, the value of which is expected to be universal, is denoted
\begin{equation}
  \Gamma_c \equiv a_{0,0} \;.
\end{equation}
The orders of the expansions are defined by 4 integers $m_1, m_2, n_1$ and $n_2$.
The quality of the fit to the data is assessed using the $\chi^2$ statistic and by calculating the goodness of fit probability.
We systematically performed fits for various orders of the expansions.
After rejecting fits for which the goodness of fit was too small, typically $Q\ll 0.1$, we chose the
fit with smallest number of parameters for which the estimation of the parameters was stable against increase in the orders of the expansions.
In each case, to determine the precision of the estimates of the fitted parameters we generated 400 synthetic data sets and determined $95\%$ confidence intervals from the fits to these synthetic data sets.
We refer the reader to Sec. 2.5 of Ref. \citen{Slevin14} for further details.
In comparison with that reference, we use slightly different definition of $w$. This change of definition can
be absorbed in a redefinition of the coefficients $b_{i,j}$ and has no effect on the fit.

We demonstrated single parameter scaling graphically as follows. First, if necessary, we subtract from the data the corrections arising from irrelevant variables, i.e.
for each data point we calculate
\begin{equation}
  \Delta\left( W, L \right) = F \left( \phi_1, \phi_2 \right) - F \left( \phi_1, 0 \right) \;,
\end{equation}
and then subtract this from the data point
\begin{equation}
  \Gamma_{\mathrm{corrected}}\left( W, L \right) = \Gamma\left( W, L \right) - \Delta\left( W, L \right) \;.
\end{equation}
We then plot the data versus $\phi_1$. In addition, we plot the curve
\begin{equation}
  F_1 \left( \phi_1 \right) = F \left( \phi_1, 0 \right) \;.
\end{equation}
on the same figure.
If the data obey single parameter scaling, all the data should collapse (within the precision of the data) onto this curve

\begin{table}[htb]
\begin{center}
\begin{tabular}{|l|l|l|l|l|l|l|l|}
  \hline
         &$W$ & $L$ & Orders of expansions & $N_{\mathrm{D}}$ & $N_{\mathrm{P}}$ & $\chi_\mathrm{min}^2$ & $Q$ \\ \hline
   3D & all & all & $m_1=2, m_2=0, n_1=2, n_2=1$ & 171  &10 & 165.5 & $\approx 0.4$ \\
         & $[18.5,19.4]$ & all & $m_1=1, m_2=0, n_1=3, n_2=1$ & 70 & 11 & 50.6 & $\approx 0.8$ \\
         & all & $12, 16, 24, 32$ & $m_1=2, n_1=2$ & 96 & 6 & 80.9 & $\approx 0.7$ \\ \hline
   4D & all & all & $m_1=2, n_1=3$ &  134 & 7 & 144.2 & $\approx 0.15$ \\
         & $[36,39]$ & all & $m_1=1, n_1=3$ &  50 & 6 & 56.4 & $\approx 0.1$ \\
         & all & $12,16,20,24$& $m_1=2, n_1=3$ &  70 & 7 & 79.1 & $\approx 0.1$ \\
  \hline
\end{tabular}
\end{center}
\caption{The range of data, the orders of the expansions, the total number of data $N_{\mathrm{D}}$, the number of parameters $N_{\mathrm{P}}$, the value of $\chi^2_{\mathrm{min}}$ obtained for the best fit, and the corresponding goodness of fit probability $Q$.}
\label{table1}
\end{table}

\begin{table}[htb]
\begin{center}
\begin{tabular}{|l|l|l|l|l|}
  \hline
    & $W_\mathrm{c}$ & $\Gamma_\mathrm{c}$ & $\nu$ & $y$ \\ \hline
  3D & $18.832[.828,.836]$ & $1.805[.803,.808]$ & $1.443[.437,.449]$ & $-3.1[-3.9,-2.4]$ \\
   & $18.835[.829,.842]$ & $1.807[.804,.812]$ & $1.433[.402,.469]$ & $-2.6[-3.7,-1.8]$ \\
   & $18.830[.827,.833]$ & $1.804[.802,.806]$ & $1.443[.437,.449]$ & \\
   \hline
  4D & $37.45[.41,.49]$ & $2.785[.771,.800]$ & $1.11[.09,.12]$ &  \\
        & $37.46[.41,.52]$ & $2.786[.768,.806]$ & $1.12[.05,.18]$ &  \\
        & $37.44[.35,.53]$ & $2.778[.730,.820]$ & $1.14[.11,.17]$ &  \\
  \hline
\end{tabular}
\end{center}
\caption{The results of the finite size scaling analyses. Details of the fits are given in the
corresponding row of Table \ref{table1}.}
\label{table2}
\end{table}

\subsection{3D}

\begin{figure}
\includegraphics[width=15cm]{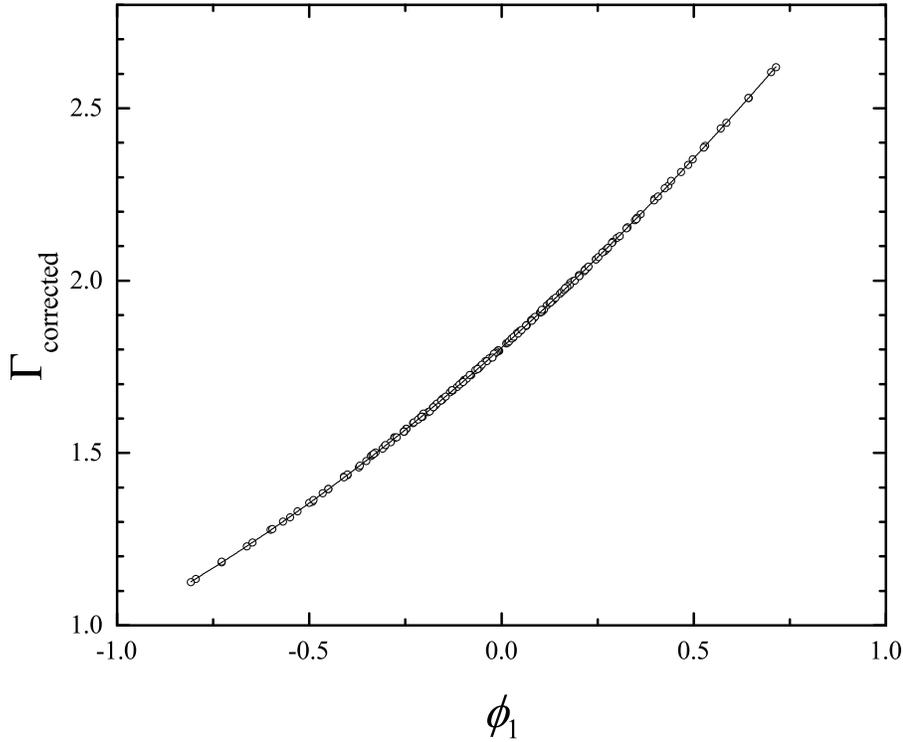}
\caption{The numerical data with the corrections to scaling subtracted (circles) are plotted versus the relevant scaling variable. The scaling function $F_1$ (sold line) is also shown. The plot demonstrates the collapse of all the data onto a single curve that is required by the single parameter scaling hypothesis.}
\label{fig3}
\end{figure}

We show the finite size scaling fit in Fig. \ref{fig1} and give the details in Tables \ref{table1} and \ref{table2}.
We found that it was not possible to fit the full data set without including a correction due to an irrelevant variable.
To check whether the inclusion of nonlinearity in the scaling variables affects the estimate of the critical parameters we narrowed the disorder range considered and fitted the data again.
The results are entirely consistent with the fit of the full data set.
To check if the inclusion of irrelevant scaling variables influences the estimation of the critical exponent we excluded smaller system sizes and fitted the data again without such corrections.
Again the results are entirely consistent with the fit of the full data set.
Full details of these checks are given in the tables.
We present the demonstration of single parameter scaling in Fig. \ref{fig3}.

\subsection{4D}
\begin{figure}
\includegraphics[width=15cm]{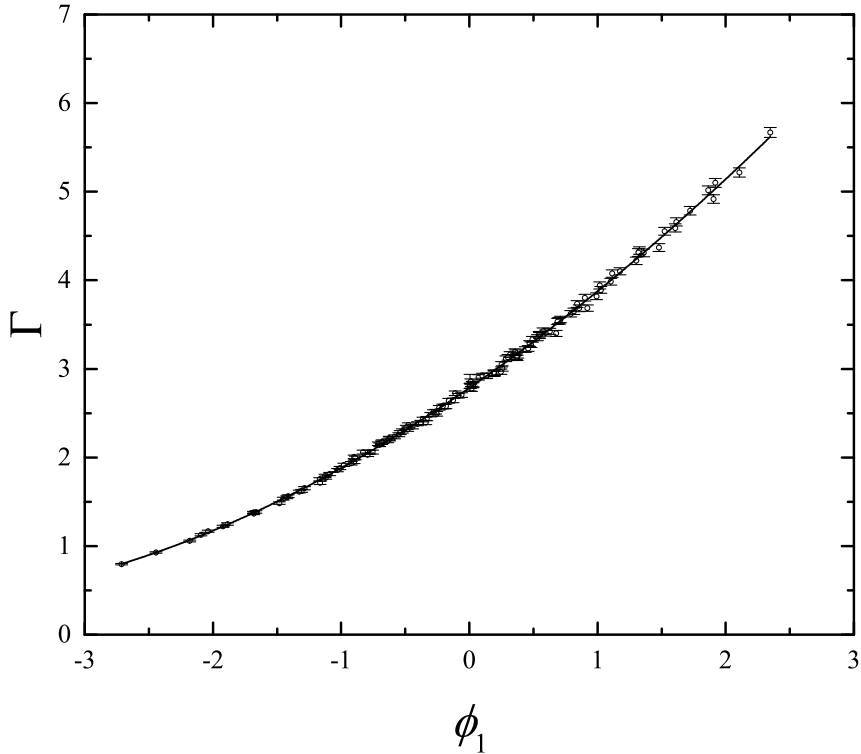}
\caption{The numerical data (circles) are plotted versus the relevant scaling variable. The scaling function $F$ (sold line) is also shown. }
\label{fig4}
\end{figure}

We show the finite size scaling fit in Fig. \ref{fig2} and give the details in Tables \ref{table1} and \ref{table2}.
We found that it was not necessary to include corrections to scaling due to irrelevant variables.
Again we also performed fits on a narrower disorder range, and also with smaller system sizes excluded. In both cases we found results that were entirely consistent with the fit of the full data set.
We present the demonstration of single parameter scaling in Fig. \ref{fig4}.

\section{Discussion}
\label{sec:discussion}
For 3D, in a previous study\cite{Slevin97} we found $\nu = 1.43 \pm .04$  for a system in a uniform magnetic field.
The estimate for the critical exponent that we  obtained here is consistent with that and also  more precise.
In that work, which was also a transfer matrix study of Lyapunov exponents, we also estimated the quantity $\Lambda_\mathrm{c}$ which is the inverse of $\Gamma_\mathrm{c}$.
Translating the result of our previous work for easier comparison, we find $\Gamma_\mathrm{c} = 1.760 \pm .004$.
There is a clear discrepancy ($\approx 3\%$) with the estimate obtained here.
This is unexpected since this value is expected to be universal, i.e. to depend only on dimensionality and symmetry class and not on the details of the model considered.
A possible explanation for this is the smaller system sizes used, and the neglect of corrections to scaling, in our  previous work.
Another possibility is that $\Gamma_\mathrm{c}$ is affected slightly by the anisotropy\cite{Delande14} introduced by
the magnetic field.
Another explanation would be a violation of universality but we think it unlikely.

We have also applied the scaling method proposed by Harada\cite{harada11,harada15} to our data.
In Harada's method the scaling function is expressed as a Gaussian process rather than a polynomial.
Since the non-linearity of the scaling variables (higher order terms in Eq. (\ref{eq:nonlinearity})) is neglected in that approach,
we restricted the range of disorder as described in Table \ref{table1}.  We have confirmed that the estimates of the critical disorder and critical exponent are consistent with the estimates based on the polynomial expansion.

Recently Ujfalusi and Varga\cite{Ujfalusi15} reported a multifractal finite size scaling analysis of wave functions obtained by large scale diagonalization of models in the three Wiger-Dyson symmetry classes in 3 dimensions.
They report $\nu = 1.424 [.407,.436]$ for a model with a uniform magnetic field in reasonable agreement with our result here.

So far we are unaware of any other reports of estimates of the critical exponent for the Anderson transition in the unitary Wigner-Dyson class in 4D.

Before concluding, we mention that the difference in the value of the exponents between different symmetry classes becomes smaller as dimension increases.  That is, $\nu\approx 1.57$  for the 3D orthogonal class\cite{Slevin14} is about 10 \% larger than the current estimate for the 3D unitary class $\nu\approx 1.44$, while the difference between the 4D orthogonal class $\nu\approx 1.156$\cite{Ueoka14} and the 4D unitary class $\nu\approx 1.11$ is less than 5\%.  This is consistent with the limit of infinite dimensions, where all the Wigner-Dyson classes are expected to show $\nu=1/2$.

\begin{acknowledgment}
This work was supported by JSPS KAKENHI Grants No. 15H03700, No. 24000013 and No. 26400393. Part of the numerical calculation has been performed on the Supercomputer System B of ISSP, The University of Tokyo.

\end{acknowledgment}

\bibliography{Slevin2016b}

\end{document}